\documentclass[letter,longauth,traditabstract]{aa}  

\usepackage{graphicx,psfig}
\usepackage{txfonts}
\begin{document}


\def\kpc{\mathrel{\rm kpc}}
\def\Msol{\mathrel{M_{\odot}}}
\def\Lsol{\mathrel{L_{\odot}}}
\def\fsub{\mathrel{f_{\rm sub}}}
\def\Mtot{\mathrel{M_{\rm tot}}}
\def\ls{\mathrel{\hbox{\rlap{\hbox{\lower4pt\hbox{$\sim$}}}\hbox{$<$}}}}
\def\gs{\mathrel{\hbox{\rlap{\hbox{\lower4pt\hbox{$\sim$}}}\hbox{$>$}}}}
\def\Msolpyr{\mathrel{\rm M_{\odot}\,yr^{-1}}}
\def\mas{\mathrel{\rm mas}}
\def\pc{\mathrel{\rm pc}}
\def\Ho{\mathrel{H_{\rm 0}}}
\def\oM{\mathrel{\Omega_{\rm M}}}
\def\oL{\mathrel{\Omega_{\rm \Lambda}}}
\def\kms{\mathrel{\rm km\,s^{-1}}}
\def\Mpc{\mathrel{\rm Mpc}}
\def\ksec{\mathrel{{\rm ksec}}}
\def\mag{\mathrel{\rm mag}}
\def\Gyr{\mathrel{\rm Gyr}}
\def\dls{\mathrel{D_{\rm LS}}}
\def\dos{\mathrel{D_{\rm OS}}}
\def\dol{\mathrel{D_{\rm OL}}}
\def\zl{\mathrel{z_{\rm L}}}
\def\zs{\mathrel{z_{\rm S}}}
\def\pix{\mathrel{\rm pixel}}
\def\um{\mathrel{\rm \mu m}}
\def\ergs{\mathrel{\rm erg\,s^{-1}}}
\def\rvir{\mathrel{\rm r_{virial}}}
\def\Om{\mathrel{\Omega_{\rm M}}}
\def\Ol{\mathrel{\Omega_{\rm \Lambda}}}
\def\Ho{\mathrel{H_0}}
\def\arcsec{\mathrel{\rm arcsec}}
\def\arcmin{\mathrel{\rm arcmin}}
\def\mJy{\mathrel{\rm mJy}}
\def\AA{\mathrel{\hbox{\rlap{\hbox{\raise6.5pt\hbox{{\hspace{0.5mm}\tiny$\circ$}}}}\hbox{\small A}}}}

\title{LoCuSS: Probing Galaxy Transformation Physics with
  Herschel\thanks{Herschel is an ESA space observatory with science
    instruments provided by European-led Principal Investigator
    consortia and with important participation from NASA.}}

   \author{
     G.\ P.\ Smith\inst{1}$\!\!$,   
     C.\ P.\ Haines\inst{1}$\!\!$,
     M.\ J.\ Pereira\inst{2}$\!\!$,
     E.\ Egami\inst{2}$\!\!$,
     S.\ M.\ Moran\inst{3}$\!\!$,
     E.\ Hardegree-Ullman\inst{4}$\!\!$,
     A.\ Babul\inst{5}$\!\!$,
     M.\ Rex\inst{2}$\!\!$,
     T.\ D.\ Rawle\inst{2}$\!\!$,
     Y.-Y.\ Zhang\inst{6}$\!\!$,
     A.\ Finoguenov\inst{7,8}$\!\!$,
     N.\ Okabe\inst{9}$\!\!$,
     A.\ J.\ R.\ Sanderson\inst{1}$\!\!$,
     A.\ C.\ Edge\inst{10}$\!\!$,
     \and
     M.\ Takada\inst{11}$\!\!$
   }

   \institute{School of Physics and Astronomy, University of
     Birmingham, Edgbaston, B15 2TT, England. \email{gps@star.sr.bham.ac.uk}
     \and
     Steward Observatory, University of Arizona, 933 North Cherry
     Avenue, Tucson, AZ85721, USA
     \and
     Department of Physics and Astronomy, The Johns Hopkins
     University, 3400 N. Charles Street, Baltimore, MD 21218, USA
     \and
     Rensselaer Polytechnic Institute (RPI) 110 Eighth Street, Troy,
     NY 12180, USA 
     \and
     Department of Physics and Astronomy, University of Victoria, 3800
     Finnerty Road, Victoria, BC, Canada 
     \and
     Argelander-Institut f\"ur Astronomie, Universit\"at Bonn, Auf
     dem H\"ugel 71, 53121 Bonn, Germany 
     \and
     Max-Planck-Institut f\"ur extraterrestrische Physik,
     Giessenbachstra\ss{}e, 85748 Garching, Germany 
     \and
     Center for Space Science Technology, University of Maryland Baltimore
     County, 1000 Hilltop Circle, Baltimore, MD 21250, USA
     \and
     Academia Sinica Institute of Astronomy and Astrophysics,
     P.O. Box 23-141, 10617 Taipei, Taiwan 
     \and
     Institute of Computational Cosmology, University of Durham, South
     Road, Durham, DH1 3LE, England
     \and
     Institute for Physics \& Mathematics of the Universe,
     University of Tokyo, 5-1-5 Kashiwa-no-Ha, Kashiwa City, 
     277-8582, Japan 
   }

   \date{Received March 31, 2010; Accepted May 19, 2010}

 
   \abstract{ We present an early broad-brush analysis of
     \emph{Herschel}/PACS observations of star-forming galaxies in 8
     galaxy clusters drawn from our survey of 30 clusters at
     $z{\simeq}0.2$.  We define a complete sample of $192$
     spectroscopically confirmed cluster members down to $L_{\rm
       TIR}{>}3{\times}10^{10}{\Lsol}$ and $L_K{>}0.25{\Lsol}$.  The
     average $K$-band and bolometric infrared luminosities of these
     galaxies both fade by a factor of ${\sim}2$ from clustercentric
     radii of ${\sim}2r_{200}$ to ${\sim}0.5r_{200}$, indicating that
     as galaxies enter the clusters ongoing star-formation stops first
     in the most massive galaxies, and that the specific
     star-formation rate (SSFR) is conserved.  On smaller scales the
     average SSFR jumps by ${\sim}25\%$, suggesting that in cluster
     cores processes including ram pressure stripping may trigger a
     final episode of star-formation that presumably exhausts the
     remaining gas.  This picture is consistent with our comparison of
     the \emph{Herschel}-detected cluster members with the cluster
     mass distributions, as measured in our previous weak-lensing
     study of these clusters.  For example, the spatial distribution
     of the \emph{Herschel} sources is positively correlated with the
     structures in the weak-lensing mass maps at ${\sim}5{\sigma}$
     significance, with the strongest signal seen at intermediate
     group-like densities.  The strong dependence of the total cluster
     IR luminosity on cluster mass -- $L_{\rm TIR}{\propto}M_{\rm
       virial}^2$ -- is also consistent with accretion of galaxies and
     groups of galaxies (i.e.\ the substructure mass function) driving
     the cluster IR luminosity.  The most surprising result is that
     roughly half of the \emph{Herschel}-detected cluster members have
     redder $S_{100}/S_{24}$ flux ratios than expected, based on the
     Rieke et al.\ models.  On average cluster members are redder than
     non-members, and the fraction of red galaxies increases towards
     the cluster centers, both of which indicate that these colors are
     not attributable to systematic photometric errors.  Our future
     goals include to intepret physically these red galaxies, and to
     exploit this unique large sample of clusters with unprecedented
     multi-wavelength observations to measure the cluster-cluster
     scatter in S0 progenitor populations, and to intepret that
     scatter in the context of the hierarchical assembly of
     clusters. }


\authorrunning{G.\ P.\ Smith, et al.}
\titlerunning{Galaxy Transformation -- A First View From Herschel}

   \maketitle

\sloppy

\section{Introduction}

Lenticular galaxies (hereafter S0s) are mainly found in the cores of
galaxy clusters at low redshift (e.g.\ Dressler et al.\ 1997; Smith et
al.\ 2005; Postman et al.\ 2005).  There is a broad consensus that
they are the descendants of gas rich spiral galaxies that have been
accreted from the surrounding filamentary structure.  However the
physics of how spirals are transformed into S0s remains largely
unconstrained, with numerous ``S0 progenitor'' populations (e.g.\
Moran et al.\ 2006; Poggianti et al.\ 2000; Geach et al.\ 2006; Haines
et al.\ 2009a -- hereafter H09a) and physical processes (e.g.\ Gunn \&
Gott 1972; Moore et al.\ 1999) discussed in the literature.

The broad range of cluster-centric radii at which various S0
progenitors are found reflects the fact that different physical
processes act in different environments, for example ram pressure
stripping is more effective closer to cluster centers where the
intracluster medium (ICM) is denser, and galaxy-galaxy merging is more
effective in galaxy groups that are falling into the cluster than in
the cluster cores.  Moreover, the observational signatures of S0
progenitors are diverse, ranging in wavelength from ultraviolet (UV)
emission from A stars in galaxies whose star-formation (SF) has been
recently quenched, through optical spectral features including Balmer
absorption lines, to mid/far-infrared (IR) emission from dust heated
by SF (e.g.\ Moran et al.\ 2007; Poggianti et al.\ 2000; Haines et
al.\ 2009b).

Mid- and far-IR properties of cluster galaxies have been studied
previously with \emph{IRAS} (e.g.\ Leggett et al.\ 1987; Doyon \&
Joseph 1989), \emph{ISO} (see Metcalfe et al.\ 2005 for a review), and
\emph{Spitzer} (e.g.\ Geach et al.\ 2006; Fadda et al.\ 2008; Haines
et al.\ 2009a,b; Bai et al.\ 2009).  A key result from these IR
studies is that a significant fraction of the total SF in galaxy
clusters is obscured by dust.  The inferred levels of SF naturally fit
the hypothesis that bulge dominated S0s are descended from late-type
spirals.  It has also been suggested that dusty S0 progenitors are
more common in dynamically active, i.e.\ merging, galaxy clusters than
in so-called ``relaxed'' clusters (e.g.\ Metcalfe et al.\ 2005; Geach
et al.\ 2006; Miller et al.\ 2006).  However it has thus far been
difficult to test this idea robustly because the intrinsic scatter in
levels of SF in clusters appears to be large, (as noted by Kodama et
al.\ 2004), and the sample sizes observed to date are small (i.e.\
$\ls2$) within any given redshift bin -- although see H09a for a
recent counter-example.

We are therefore conducting a systematic wide-field survey of a large
statistically well-defined sample of galaxy clusters in a narrow
redshift slice at $z{\simeq}0.2$, as part of the Local Cluster
Substructure Survey
(LoCuSS\footnote{http://www.sr.bham.ac.uk/locuss/}).  Our goals are to
compile a complete inventory of S0 progenitors using data from the
far-UV to far-IR, and to relate these populations to the underlying
gas physics and hierarchical structure of the host galaxy clusters.
We aim to delineate the different physical processes responsible for
galaxy transformation in clusters and their surrounding large scale
structure, and thus constrain the amplitude of the different physical
pathways from spiral to S0 morphology, and how these relate to the
dynamical state of the clusters.  Our Open Time Key Programme
observations with \emph{Herschel} (Pilbratt et al., 2010),
supplemented by existing \emph{Spitzer} mid-IR observations provide
the all-important measurements of the bolometric IR luminosity and
mid/far-IR colors of dust-reddened/obscured S0 progenitors.

We assume ${\Ho}{=}70{\kms}$, ${\Om}{=}0.3$, ${\Ol}{=}0.7$.  In this
cosmology $1{\kpc}$ at $z{=}0.2$, subtends $0.3''$.  All cluster
masses and radii relative to an over-density are derived from the
weak-lensing analysis of Okabe et al.\ (2010; hereafter Ok10).

\section{Survey Design}

Our survey goals include understanding the physical reasons for the
large cluster-cluster variations in SF rate (SFR), and the full range
of physical processes responsible for transforming spiral galaxies
into S0s. We therefore require a large sample of clusters in order to
sample thoroughly the underlying cluster population and the various S0
progenitor populations that they host.  Our sample of 30 clusters
therefore will allow us to study, for example, how total integrated
cluster SFRs depend on global cluster properties such as cool core
strength, and substructure fraction, in ${\sim}3{-}6$ cluster bins with
${\sim}5{-}10$ clusters per bin.  Based on previous IR and UV studies
of S0 progenitors, we expect ${\sim}30{-}50$ such objects per cluster.
Our sample of 30 clusters should therefore deliver a sample of
${\gs}1000$ S0 progenitors.

With current observing facilities and observed cluster samples, this
study is prohibitvely expensive at high redshift because of the
requirement for wide-field and moderately deep data on a large sample.
We therefore concentrate on clusters at $z{\simeq}0.2$; at this
redshift gravitational lensing is an efficient probe of the dark
matter distribution in the cluster in-fall regions using Suprime-CAM
on the Subaru telescope, and yet follow-up \emph{Herschel}
observations of a sample of 30 clusters are feasible.

The cluster sample is a subset of those in the \emph{ROSAT} All-sky
Survey catalogs (Ebeling et al.\ 1998, 2000; B\"ohringer et al.\ 2004)
that satisfy the following criteria: $0.15{<}z{<}0.3$,
$n_H{<}7{\times}10^{20}{\rm cm}^{-2}$, and
${-}70^\circ{<}\delta{<}{+}70^\circ$, and that were observable with
Subaru/Suprime-CAM on the nights assigned to us (Ok10).  The sample is
therefore blind to the thermodynamic, and hierarchical assembly
history of the clusters, other than the use of X-ray luminosity as a
proxy for mass-selection.  The distribution of the X-ray luminosities
of clusters in our sample is statistically indistinguishable from that
of a volume-limited sample satisfying
$L_X\,E(z)^{-2.7}{>}4.2{\times}10^{44}{\ergs}$, where the scaling of
$L_X$ with $E(z)^2{=}\Omega_{\rm M}\,(1{+}z)^3{+}\Omega_{\Lambda}$
approximates mass-selection following Popesso et al.\ (2005) -- see
Ok10 for more details.

Our multi-wavelength observations of this sample (\S3) span at least a
clustercentric radius of $12.5{\arcmin}$ on the sky, equating to
${\sim}1.5r_{200}$ for a typical cluster in our sample.  This physical
field of view is sufficiently large to probe all of the physical
processes expected to play a role in galaxy transformation (Fig.~1).

\begin{figure}
  \centerline{
    \psfig{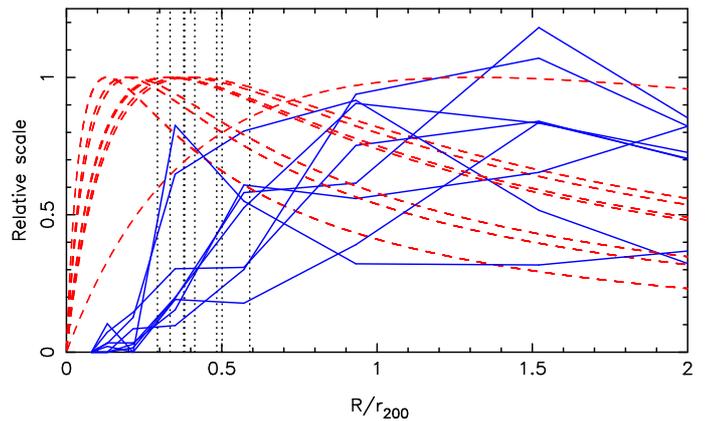}
  }
  \caption{This figure shows simple physically well-motivated models
    for the relative strength of ram pressure stripping, harassment,
    and galaxy-galaxy merging in the eight clusters discussed in this
    letter.  The vertical black dotted lines show the stripping radius
    for each cluster -- the radius within which ram pressure is able
    to strip a disk galaxy with rotation speed of $160\kms$.  This
    calculation combines Sanderson et al.'s (2009)
    \emph{Chandra}-based gas density profiles with Ok10's weak
    lensing-constrained NFW density profiles, and assumes that the
    typical radial velocity of infalling galaxies is $0.9V_C$ at the
    virial radius where $V_C$ is the circular velocity of the cluster
    dark matter halo, following the analytic model of McCarthy et al.\
    (2008).  Red dashed curves show the relative efficiency of
    harassment, based on the harassment rate of $f_{\rm
      H}\propto\rho_{\rm gal}r^2$ (Moore et al.\ 1998), assuming that
    the total mass distribution traces the galaxy density (Lin et al.\
    2004) -- i.e.\ we substitute $\rho_{\rm NFW}$ from Ok10 for
    $\rho_{\rm gal}$.  The galaxy-galaxy merger rate is traced by the
    solid blue curves, which show the radial number density profile of
    galaxies that inhabit group-like environments, i.e.\ a
    3-dimensional galaxy density of $2{\le}{\rho}_{\rm
      3D}{\le}10{\Mpc}^{-3}$, derived from our near-IR photometry
    (H09a) and large spectroscopic catalogs (Hardegree-Ullman et al.\
    in prep.).}

\end{figure}

\section{Observations and Data Analysis}

The 8 clusters discussed in this letter were observed with the
Photodetector Array Camera and Spectrometer (PACS; Poglitsch et
al. 2010) across a $25'{\times}25'$ field of view at $100$ and
$160{\um}$ in scan map mode at $20{\arcsec}/{\rm s}$ in November and
December 2009.  Each cross scan was repeated 7 times, giving a total
exposure time of $93.2{\rm s/pix}$.  We first processed the data using
standard {\sc hipe} routines (Ott 2010).  Then all sources detected
at ${\ge}2.5{\sigma}$ in the first pass reduced data were masked using
$15''$ circular apertures, and the data were high pass filtered with a
filter 25 and 30 frames wide at $100{\um}$ and $160{\um}$
respectively.  The final maps were then constructed using the {\sc
  photproject} routine.  The angular resolution of the final reduced
frames is $6.8''$ at $100{\um}$ and $11.4''$ at $160{\um}$.

Sources were extracted using SExtractor, employing circular apertures
of $12''$ and $16''$ diameter at $100$ and $160{\um}$.  The point
spread function in the $100{\um}$ frame is slightly de-graded from
that derived from the PACs calibration maps of Vesta, due to imperfect
spatial calibration within scans where bright sources cannot be used
to register the individual exposures.  We therefore applied an
empirical aperture correction of $1.65{\pm}0.03$, based on brightest
isolated sources in the $100{\um}$ maps.  The standard Vesta aperture
correction of $1.675$ was applied at $160{\um}$.  The $90\%$
completeness limits are $18{\mJy}$ at $100{\um}$ and $28{\mJy}$ at
$160{\um}$.  Calculation of the total IR luminosities
($8{<}{\lambda}{<}1000{\um}$) discussed in \S4 are described by Haines
et al.\ (2010).

We also use our wide-field data from \emph{Spitzer}/MIPS ($24{\um}$;
H09a), UKIRT/WFCAM ($J/K$-bands; H09a), \emph{Chandra} (Sanderson et
al., 2009), and Subaru/Suprime-CAM ($V/i'$-bands; Okabe \& Umetsu
2008; Ok10).  We have spectroscopically identified $92\%$ of the
sources down to the $100\um$ detection threshold using MMT/Hectospec
(Fabricant et al.\ 2005), in addition to securing ${\sim}200{-}300$
cluster galaxy redshifts per cluster (Hardegree-Ullman et al., in
prep.).

\section{Results}

\begin{figure}
  \centerline{
    \psfig{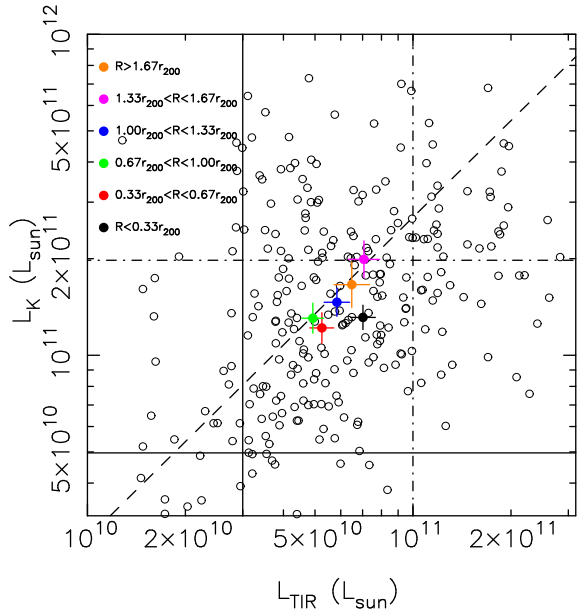}
  }
  \caption{$K$-band luminosity ($L_{\rm K}$) versus total infrared
    luminosity ($L_{\rm TIR}$) for the \emph{Herschel}-detected
    cluster members, plotted as open circles.  The horizontal and
    vertical solid lines show the completeness limits of our
    MMT/Hectospec spectroscopic survey and our \emph{Herschel}
    photometry respectively; the horizontal and vertical dot-dashed
    lines show $L_K^\star$ from Lin et al.\ (2004) and the definition
    of LIRGs respectively.  The diagonal dashed line shows $L_{\rm
      K}{\propto}L_{\rm TIR}$, and guides the eye for the radial
    dependence of ${\langle}L_K{\rangle}$ and ${\langle}L_{\rm
      TIR}{\rangle}$ respectively, as shown by the color-coded filled
    circles -- see legend.  }
\end{figure}

We identify $192$ \emph{Herschel} sources down to $L_{\rm
  TIR}{>}3{\times}10^{10}L_\odot$ and $L_K{>}0.25L_K^\star$ (Fig.~2)
within a clustercentric radius of $R{<}1.5r_{200}$ and lying inside
the caustics in the velocity-radius plane (see Haines et al.\ 2010 for
an example).  Just $(24{\pm}3)\%$ (46/192) of these galaxies are LIRGs
($L_{\rm TIR}{>}10^{11}L_\odot$) and none are ultra-luminous IR
galaxies (ULIRGs; $L_{\rm TIR}{>}10^{12}L_\odot$).  The most luminous
galaxy is the brightest cluster galaxy (BCG) in A\,1835; we also
detect the BCG in A\,2390 (Edge et al.\ 1999; Egami et al.\ 2006).
The typical galaxy has $L_K{\simeq}1.5{\times}10^{11}L_\odot$ and
$L_{\rm TIR}{\simeq}6{\times}10^{10}L_\odot$.  On average, the most
IR-luminous galaxies are found at projected cluster-centric radii of
${\sim}1.5r_{200}$; at larger and smaller radii the average
IR-luminosity declines.  This trend is mirrored by a decline in the
average $K$-band luminosity such that ${\langle}L_{\rm
  TIR}/L_K{\rangle}$ is conserved down to ${\sim}0.5r_{200}$, interior
to which the average IR-luminous galaxy jumps from ${\langle}L_{\rm
  TIR}/L_K{\rangle}(R{\gs}0.5r_{200}){\simeq}0.39{\pm}0.03$ to
${\langle}L_{\rm
  TIR}/L_K{\rangle}(R{\ls}0.5r_{200}){\simeq}0.53{\pm}0.03$ (Fig.~2).

\begin{figure*}
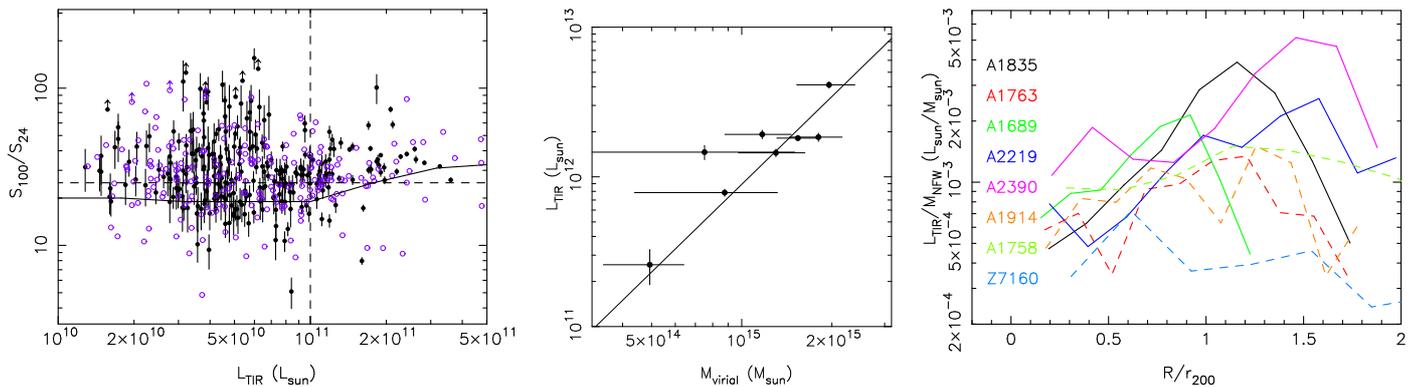

  \centerline{
    \psfig{file=collum_label.ps,height=50mm,angle=0}
    \hspace{2mm}
    \psfig{file=lirmass_20100517_label.ps,height=50mm,angle=0}
    \hspace{2mm}
    \psfig{file=profile_20100517_200_label.ps,height=50mm,angle=0}
  }
  \caption{{\bf Left} -- Far-IR color-luminosity relation; the
    color-luminosity relation predicted by the Rieke et al.\ (2009)
    SED templates is shown as the black solid line.  The vertical and
    horizontal dashed lines mark the definition of LIRGs, and the
    nominal color cut discussed in \S4, respectively.  Cluster members
    are shown as filled black circles, and non-members as open purple
    circles.  {\bf Center} -- Cluster virial mass versus bolometric
    infrared luminosity within the virial radius; the best-fit
    relation is shown as a solid line.  {\bf Right} -- Total IR
    luminosity density profile relative to total mass density profile
    (\S4).  Strong-lensing clusters (Richard et al.\ 2010) are plotted
    as solid curves and non-strong-lensing clusters as dashed curves.
    The legend lists clusters from most (A\,1835) to least (Z\,7160)
    massive.  }
\end{figure*}

The $S_{100}/S_{24}$ colors of the IR-detected galaxies have a
prominent excess of flux at $100{\um}$ relative to that expected from
the commonly used Rieke et al.\ (2009) SED templates (Fig.~3).
Adopting $S_{100}/S_{24}{>}25$ as defining this unexpected red
population, we find that $(63{\pm}4)\%$ (121/192) of the
spectroscopically confirmed members are ``red'', with
$(78^{+6}_{-8})\%$ (36/46) of LIRG members, and $(58^{+4}_{-5})\%$
(85/146) of sub-LIRG members being ``red'' respectively.  The
predominance of red LIRGs is partly expected given the predicted
relationship between luminosity and color, however the observed LIRGs
lie almost exclusively red-ward of the Rieke et al.\ models.  The
fraction of galaxies with red colors also shows a gentle increase
towards the cluster centers: $f_{\rm red}{\propto}R^{(-0.2{\pm}0.1)}$.
The mean color of \emph{Herschel}-detected cluster members,
${\langle}S_{100}/S_{24}{\rangle}{=}29.2{\pm}0.1$, is also slightly
redder than the mean color of \emph{Herschel}-detected non-members
(defined as lying outside the caustics, but within $0.15{<}z{<}0.3$),
${\langle}S_{100}/S_{24}{\rangle}{=}27.0{\pm}0.1$.  Both of these
correlations point to a physical origin for these red colors, rather
than systematic photometric errors.  Indeed, Rawle et al.\ (2010) find
a similar population in the core of the Bullet cluster.  However the
small difference between the color distributions of members and
non-members suggest that cluster physics may not be the main
determinant of the IR colors (see also Pereira et al.\ 2010).

\begin{figure*}
  \begin{minipage}{110mm}
  \centerline{
    \psfig{file=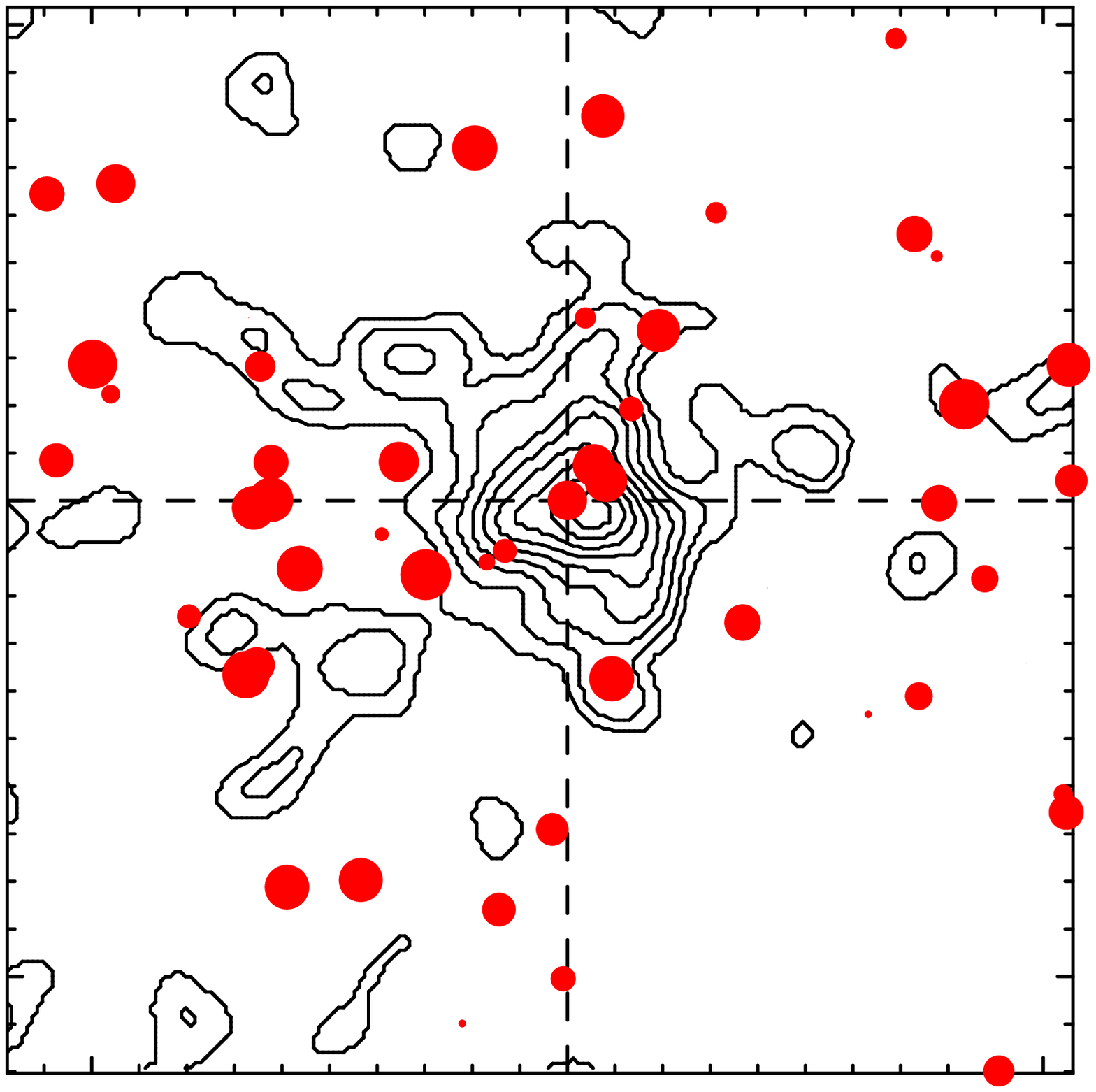,width=27.5mm,angle=0}
    \psfig{file=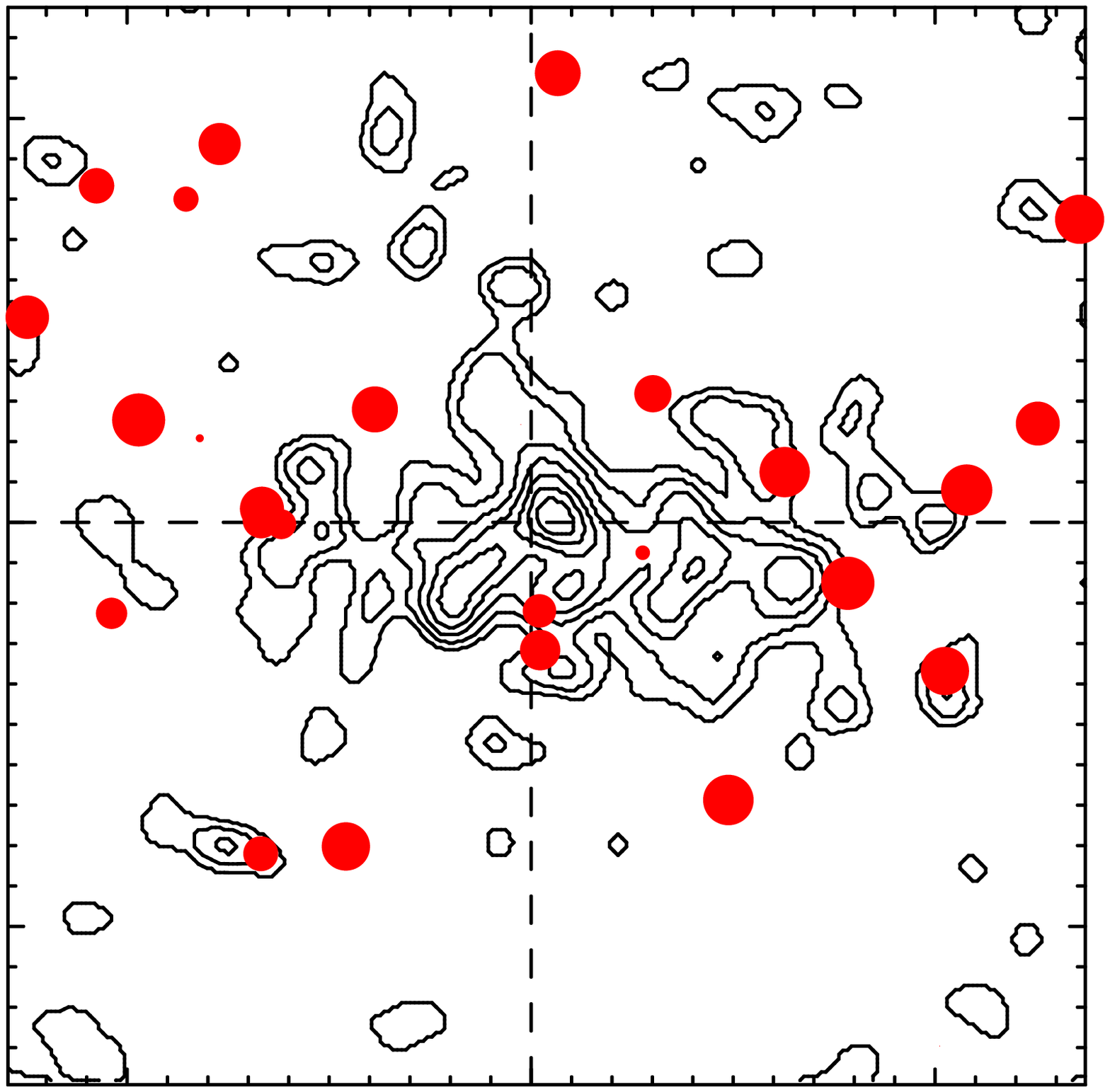,width=27.5mm,angle=0}
    \psfig{file=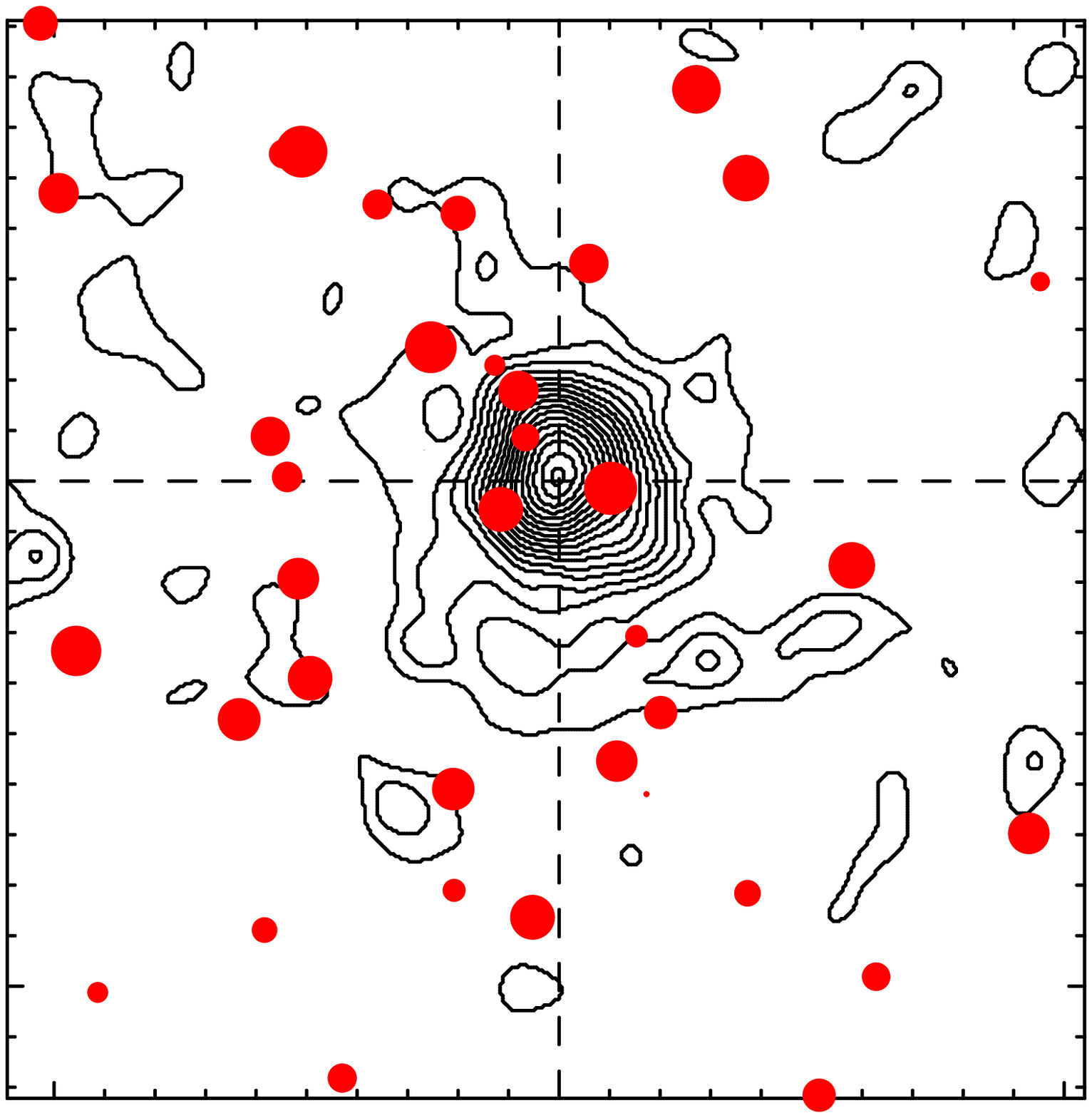,width=27.5mm,angle=0}
    \psfig{file=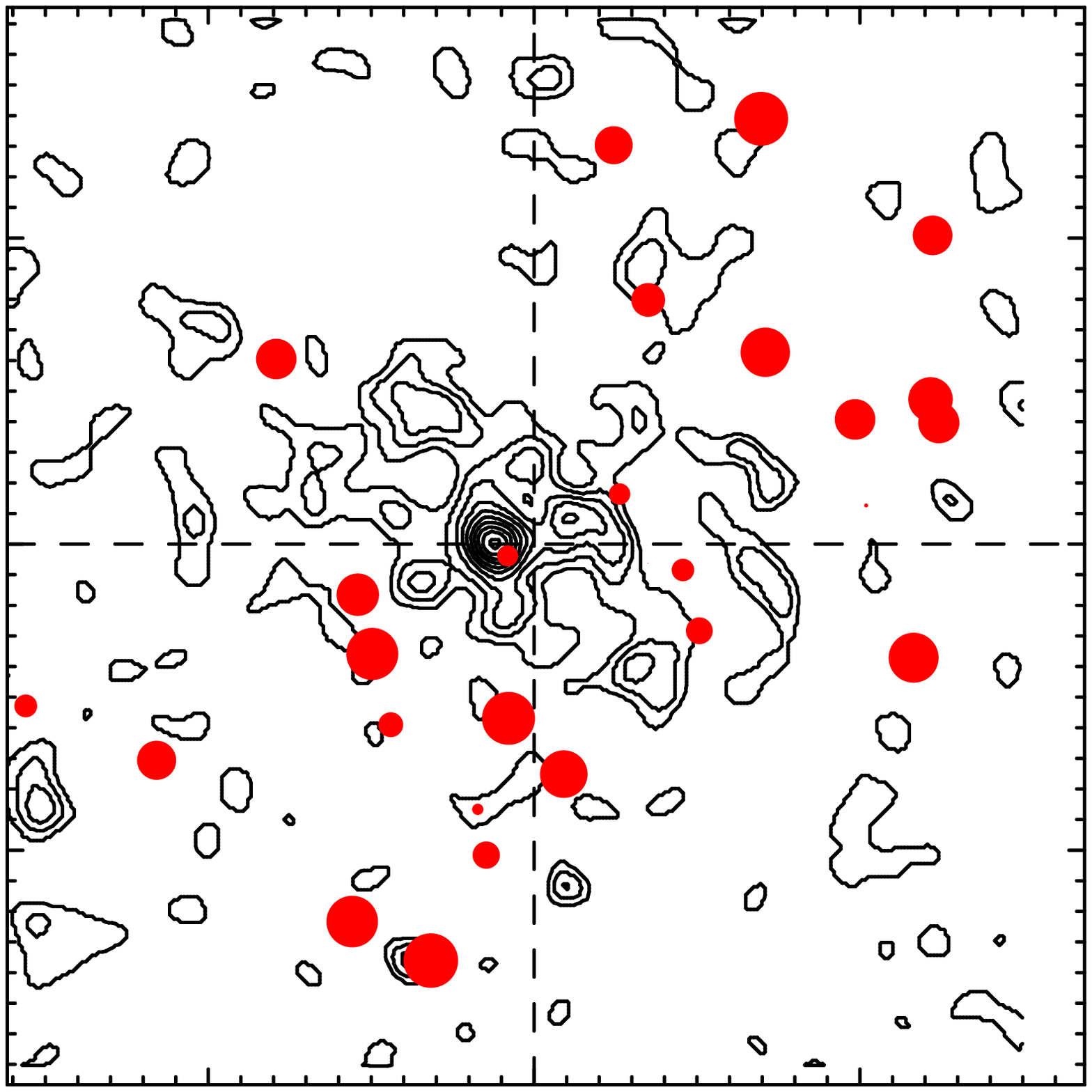,width=27.5mm,angle=0}
  }
  \centerline{
    \psfig{file=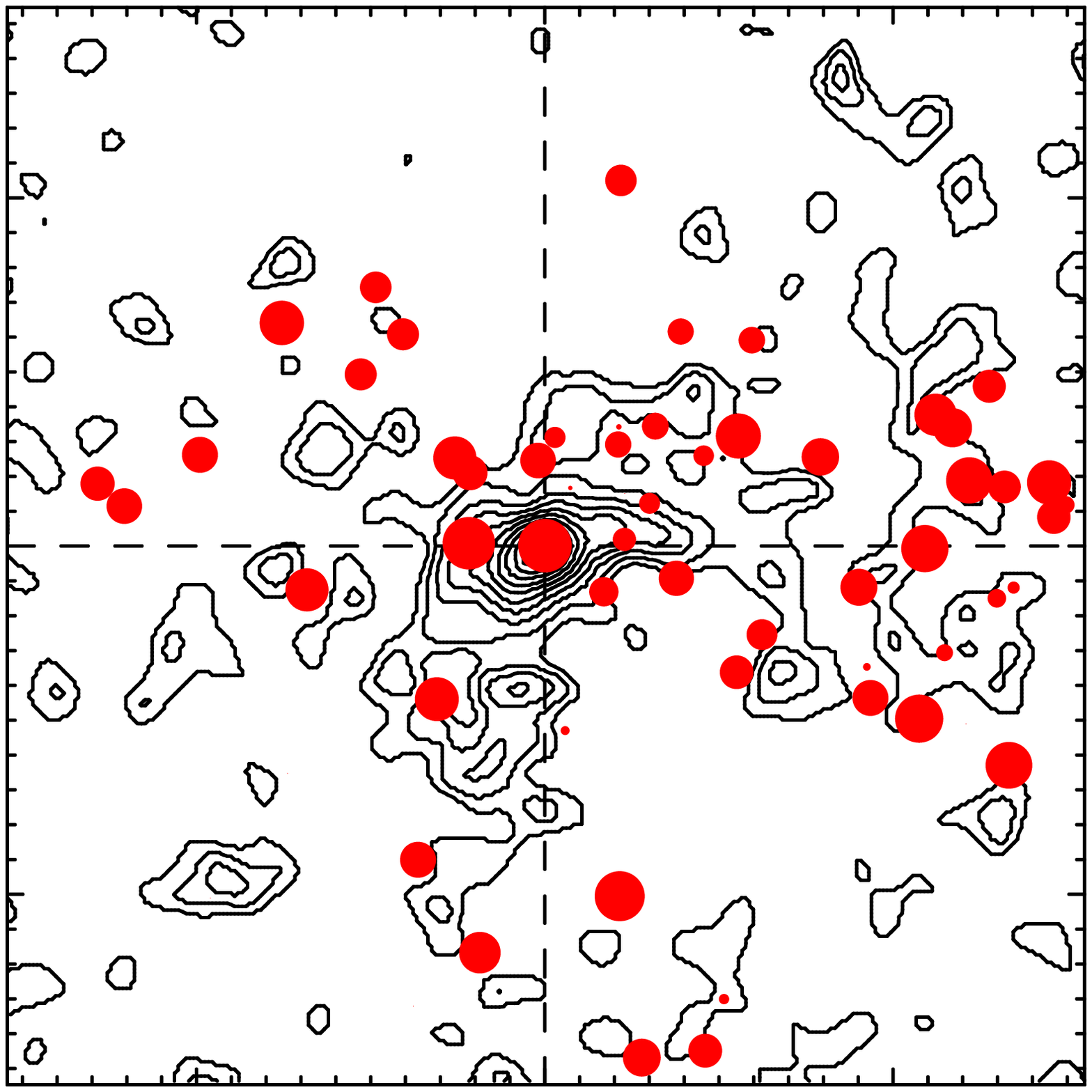,width=27.5mm,angle=0}
    \psfig{file=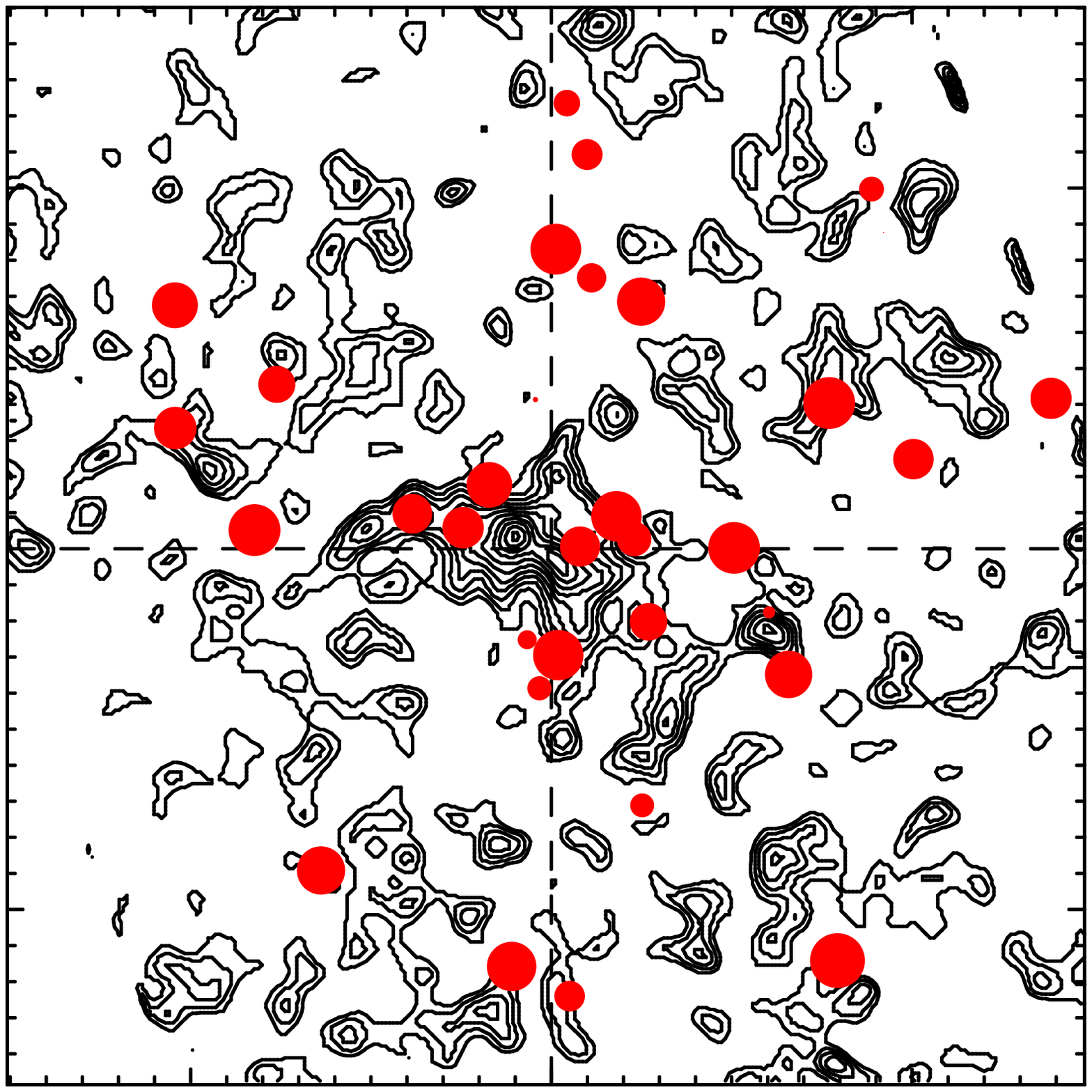,width=27.5mm,angle=0}
    \psfig{file=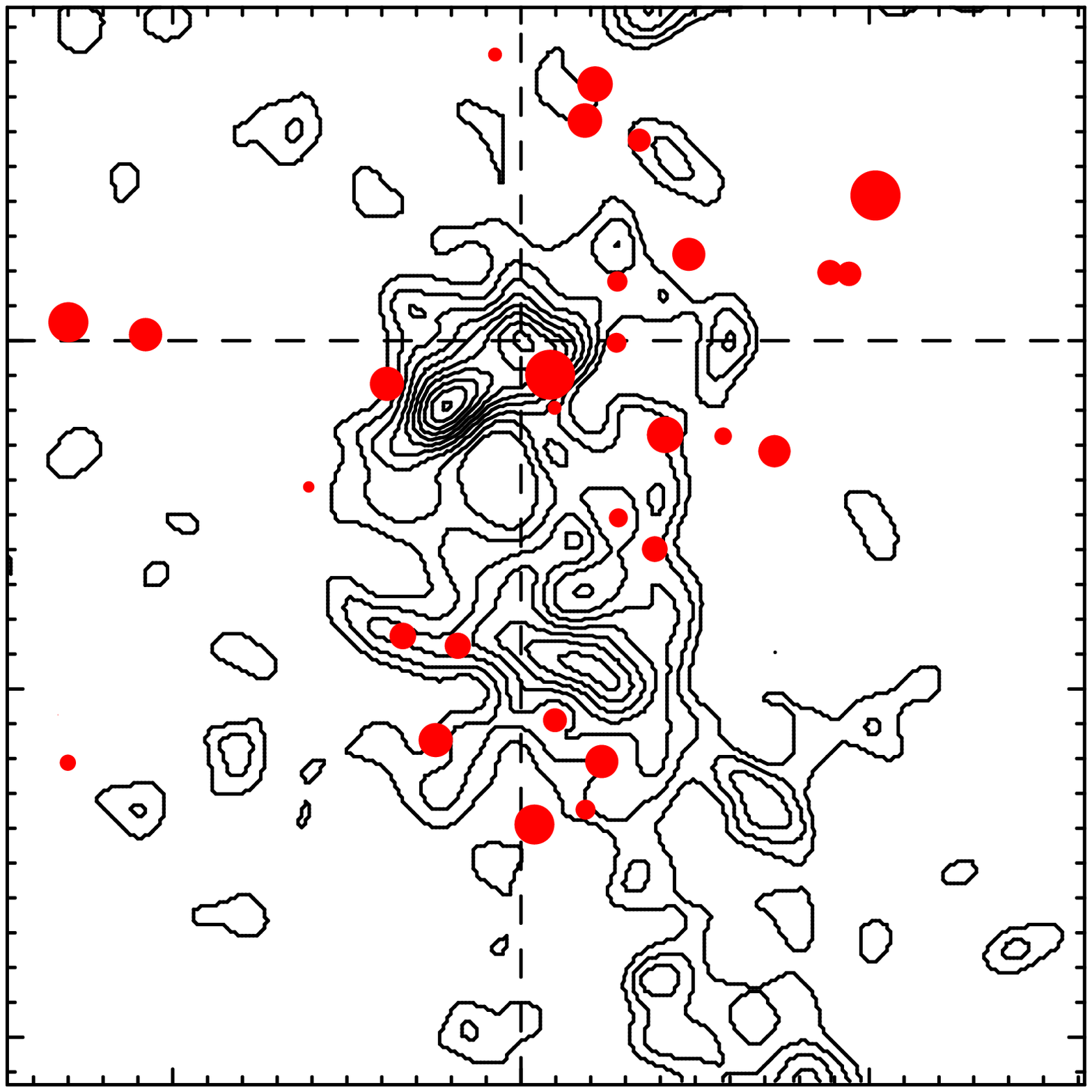,width=27.5mm,angle=0}
    \psfig{file=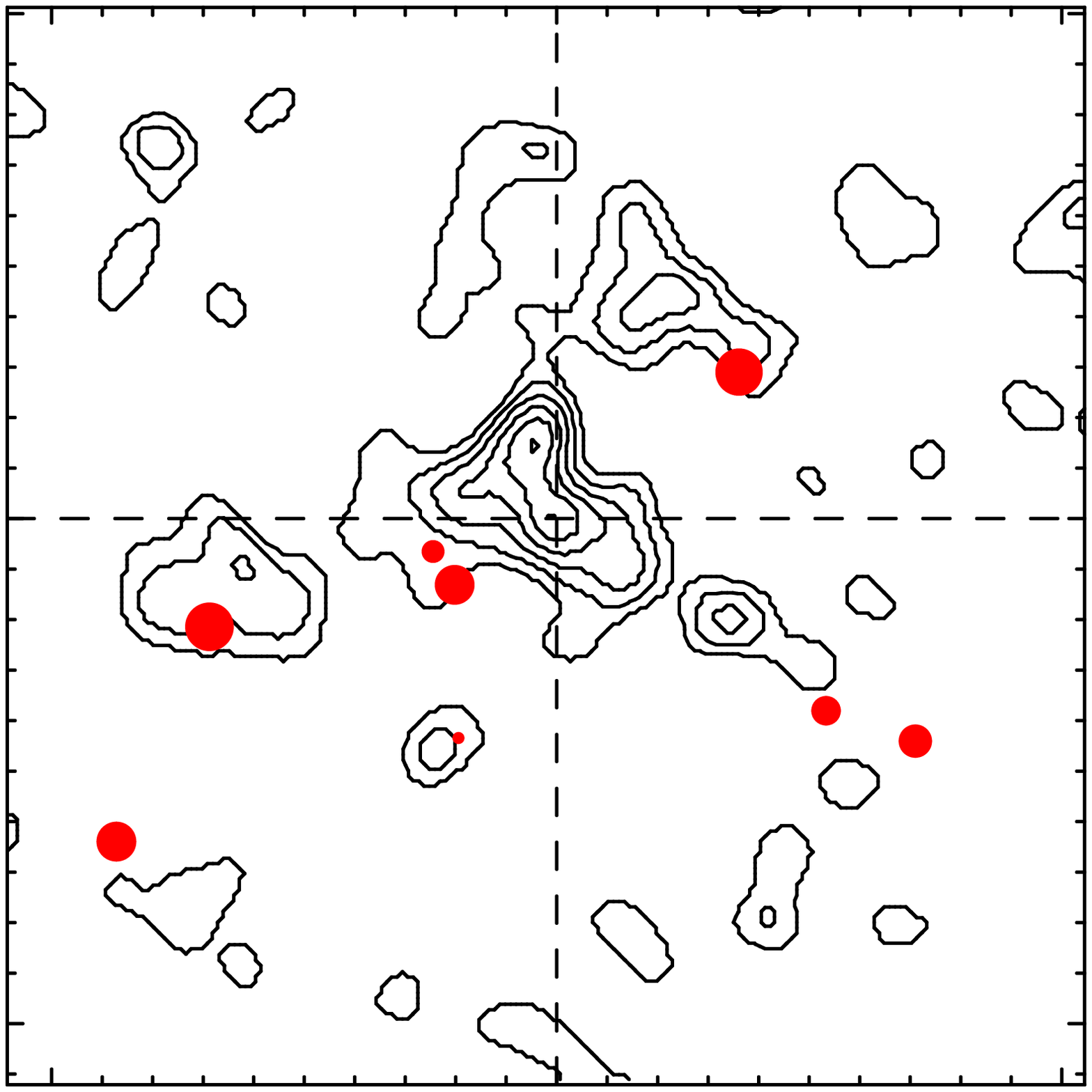,width=27.5mm,angle=0}
  }
  \end{minipage}
  \hspace{1mm}
  \begin{minipage}{80mm}
    \psfig{file=filament_20100517_label.ps,height=55mm,angle=-90}
  \end{minipage}
  \caption{{\bf Left} -- Weak lensing mass maps (contours, spaced at
    $1\sigma$ mass reconstruction error; Ok10; Okabe \& Umetsu 2008)
    in order of decreasing cluster virial mass -- top left-to-right:
    A\,1835, A\,1763, A\,1689, A\,2219; bottom: A\,2390, A\,1914,
    A\,1758, Z\,7160.  Spectroscopically confirmed
    \emph{Herschel}-detected members (selected as described in \S4)
    are marked as filled red circles that scale logarithmically with
    $L_{\rm TIR}/L_K$.  North is up and East is left, and the axis
    tick-marks are spaced at $1\arcmin$ intervals.  {\bf Right} --
    Number density of galaxies relative to the mean number density of
    each population as a function of total projected cluster mass
    density.  Positive correlation of the \emph{Herschel}-detected
    cluster members with the mass distribution in/surrounding the
    clusters is detected at ${\sim}5{\sigma}$.  The absence of red
    triangular data points at ${\Sigma}_{\rm
      mass}{>}1000{\Msol}{\kpc}^{-2}$ is because we do not find any
    non-BCG IR luminous cluster members at such high densities.  The
    dip in the signal for cluster members at ${\sim}1250{\Msol}{\rm
      kpc}^{-2}$ is due to the incompleteness of our spectroscopic
    catalog in high density environments due to MMT/Hectospec fiber
    collisions. }
\end{figure*}

The IR luminous population traces the structure of the mass
distribution of the clusters and the surrounding filamentary
structure, as traced by Ok10's weak-lensing mass maps (Fig.~4).  This
is quantified using the number density of galaxies (normalized to the
mean galaxy number density in each case) as a function of projected
cluster mass density obtained from the mass maps (Fig.~4).  A
spatially random distribution of galaxies is consistent with unity;
positive correlation between galaxies and mass is greater than unity,
and anti-correlation is less than unity.  We detect correlation
between the spatial distribution of the IR luminous cluster members
and the cluster mass distributions at ${\sim}5{\sigma}$ significance.

We also combine the \emph{Herschel} data with Ok10's weak-lensing
analysis to construct the first ever mass-$L_{\rm TIR}$ relation for
galaxy clusters (Fig.~3), obtaining $L_{\rm TIR}({<}r_{\rm
  virial}){\propto}M_{\rm virial}^\alpha$, with
${\alpha}{=}1.9^{+1.4}_{-0.5}$.  The fit was done taking into account
errors in both variables, and was repeated $10^4$ times, each time
drawing 8 clusters at random with replacement; the error quoted on the
slope is dominated by the scatter between these bootstrap samples.
The mass-to-light ratio of a ${\sim}10^{15}{\Msol}$ cluster is
${\sim}10^3{\Msol}/{\Lsol}$, however the scaling of mass with far-IR
luminosity is inconsistent with a constant mass-to-light ratio at
${\sim}2{\sigma}$.

Cluster IR luminosity density profiles, normalized to the underlying
dark matter density profile from Ok10, are generally increasing
functions of clustercentric radius out to at least $r_{200}$ (Fig.~3).
IR luminosity in excess of that expected from a flat mass-to-light
ratio profile is therefore found at large clustercentric radii,
particularly at $R{\sim}1{-}1.5r_{\rm virial}$, where some of the
profiles show a pronounced peak.  We also note that strong-lensing
clusters (A\,1835, A\,1689, A\,2219, A\,2390) tend to have steeper
mass-normalized luminosity density profiles than non-strong-lensing
clusters, three of which are well-known merging clusters (A\,1763,
A\,1758, A\,1914) in which the merger axis is likely close to the
plane of the sky.  Cluster geometry, e.g.\ prolate shape and/or merger
aligned with the line of sight (strong-lensing clusters) versus
aligned in the plane of the sky, therefore may complicate the
detailed interpretation of the luminosity density profile shapes.

\section{Summary and Discussion}

We have presented an initial broad-brush analysis of
\emph{Herschel}/PACS observations of $25\%$ of our sample of 30 galaxy
clusters at $z{\simeq}0.2$, and combined these data with our existing
\emph{Spitzer}, Subaru, \emph{Chandra}, UKIRT, and MMT data.  The main
analysis concentrates on a sample of 192 spectroscopically confirmed
cluster members with $L_{\rm TIR}{>}3{\times}10^{10}L_\odot$,
$L_K{>}0.25L_K^\star$, $R{<}1.5r_{200}$.  The average $K$-band
luminosity of these galaxies fades by a factor of almost 2 from the
cluster outskirts (${\sim}1{-}2r_{200}$) to the cluster cores
(${\ls}0.5r_{200}$), although the average specific star-formation
rate, as probed by ${\langle}L_{\rm
  TIR}{\rangle}/{\langle}L_K{\rangle}$, is constant across most of
this radial range (${\sim}0.5{-}2r_{200}$), before jumping by 25\% on
smaller scales.  This suggests that as gas rich galaxies fall into the
clusters (typically in groups -- Fig.~1) ongoing star-formation stops
first in the most massive galaxies.  As galaxies reach the cluster
cores physical processes that operate in high density environments,
for example ram pressure stripping and harrassment, then appear to
trigger a final episode of star-formation that presumably exhausts the
remaining gas supply.

This picture is consistent with our comparison of the
\emph{Herschel}-detected cluster members with the cluster mass
distributions, as probed by Okabe et al.'s (2010) weak-lensing
analysis.  First, the spatial distribution of the \emph{Herschel}
sources is positively correlated with the structures in the
weak-lensing mass maps at ${\sim}5{\sigma}$ significance, with the
strongest signal seen at intermediate, group-like densities.  Second,
the strong dependence of the total cluster IR luminosity on cluster
mass ($L_{\rm TIR}{\propto}M_{\rm virial}^2$) is consistent with
accretion of galaxies and groups of galaxies driving the cluster IR
luminosity.  This is because, assuming that IR galaxy mass-to-light
ratios are independent of the cluster mass, the scaling relation can
be understood as stemming from the $M^2$ dependence of the
substructure mass function seen in theoretical models (e.g.\ Taylor \&
Babul 2005).  Third, the IR luminosity density profiles of the
clusters generally increase to large radii, with some clusters showing
a peak at ${\sim}r_{200}$.  This is qualitatively consistent with the
enhanced star-formation rates seen in in-falling galaxy populations by
Moran et al.\ (2005).

The most surprising result is that roughly half of the
\emph{Herschel}-detected cluster galaxies have excess flux at
$100{\um}$ over that predicted from current SED models.  Cluster
members are redder than non-members, and we find a shallow trend of
increasing fraction of red IR galaxies towards the cluster centers,
both of which suggest that this is a physical effect and not caused by
systematic photometric uncertainties.  This result will be the focus
of more detailed future investigation.

Finally, we note that, contrary to previous speculation in the
literature, we do not find a strong relationship between cluster IR
luminosity and cluster dynamical state.  Observations of the full
sample will allow us to investigate this issue in more detail.

\begin{acknowledgements}
  We acknowledge the anonymous referee for helping us to clarify
  various aspects of this letter.  We thank our colleagues within the
  LoCuSS collaboration for many stimulating discussions, and their
  enthusiastic support.  GPS is supported by the Royal Society.  CPH
  and AJRS thank STFC for some support.  Support for this work was
  provided by NASA through an award issued by JPL/Caltech.  YYZ is
  supported by the German BMBF through the Verbundforschung under
  grant 50 OR 1005.
\end{acknowledgements}


\begin{thebibliography}{}
\bibitem[2009]{bai} Bai et al., 2009, ApJ, 693, 1840
\bibitem[2004]{boehringer} B\"ohringer et al., 2004, A\&A, 425, 367
\bibitem[1989]{doyon} Doyon \& Joseph, 1989, MNRAS, 239, 347
\bibitem[1997]{dressler} Dressler A., et al., 1997, ApJ, 490, 577
\bibitem[1998]{ebeling98} Ebeling et al., 1998, MNRAS,  301, 881
\bibitem[2000]{ebeling00} Ebeling et al., 2000, MNRAS, 318, 333
\bibitem[1999]{edge} Edge A.C., et al., 1999, MNRAS, 306, 599
\bibitem[2006]{egami} Egami E., et al., 2006, ApJ, 647, 922
\bibitem[2005]{fabricant} Fabricant D., et al., 2005, PASP, 117, 1411
\bibitem[2008]{fadda} Fadda D., et al., 2008, ApJ, 672, 9
\bibitem[2006]{geach} Geach J., et al., 2006, ApJ, 649, 661
\bibitem[1972]{gunngott} Gunn \& Gott, 1972, ApJ, 176, 1
\bibitem[2009a]{haines} Haines C.\ P., et al., 2009, MNRAS, 396, 1297
\bibitem[2009b]{haines} Haines C.\ P., et al., 2009, ApJ, 704, 126
\bibitem[2010]{haines} Haines C.\ P., et al., 2010, this volume
\bibitem[1987]{leggett} Leggett S., et al., 1987, MNRAS, 228, 11
\bibitem[2004]{kodama} Kodama T., et al., 2004, MNRAS, 354, 1103
\bibitem[2010]{lin} Lin Y., et al., 2004, ApJ, 610, 745
\bibitem[2008]{mccarthy} McCarthy I., et al., 2008, MNRAS, 383, 593
\bibitem[2005]{metcalfe} Metcalfe, et al., 2005, SSRv, 119, 425
\bibitem[2004]{miller} Miller et al., 2006, AJ, 131, 2426
\bibitem[1998]{moore} Moore B., et al., 1999, MNRAS, 304, 465
\bibitem[2005]{moran05} Moran S.\ M., et al., 2005, ApJ, 634, 977
\bibitem[2006]{moran06} Moran S.\ M., et al., 2006, ApJ, 641, 97
\bibitem[2007]{moran07} Moran S.\ M., et al., 2007, ApJ, 671, 1503
\bibitem[2008]{okabe08} Okabe N.\ \& Umetsu K., 2008, PASJ, 
\bibitem[2010]{okabe10} Okabe N., et al., 2010, arXiv:0903.1103
\bibitem[2010]{ott} Ott, S., 2010, in ASP Conference Series,
  Astronomical Data Analysis Software and Systems XIX, Y.\ Mizumoto,
  K.-I.\ Morita, and M.\ Ohishi, eds., in press
\bibitem[2010]{pereira} Pereira M.\ J., et al., 2010, this volume
\bibitem[2010]{pilbratt} Pilbratt et al., 2010, this volume
\bibitem[2010]{poggianti} Poggianti B.\ \& Wu, 2000, ApJ, 529, 157
\bibitem[2010]{pogtlitsch} Pogtlitsch, et al., 2010, this volume
\bibitem[2005]{popesso} Popesso, P., et al., 2005, A\&A, 433, 431
\bibitem[2005]{postman} Postman M., et al., 2005, ApJ, 623, 721
\bibitem[2010]{rawle} Rawle T.\ D., et al., 2010, this volume
\bibitem[2010]{richard} Richard J., et al., 2010, MNRAS, 404, 325
\bibitem[2009]{Rieke} Rieke G. H., et al., 2009, ApJ, 692, 556
\bibitem[2009]{sanderson} Sanderson A.\ J.\ R., et al., 2009, MNRAS, 398, 1698
\bibitem[2005]{smith05} Smith G.\ P., et al., 2005, ApJ, 620, 78
\bibitem[2005]{Taylor} Taylor \& Babul, 2005, MNRAS, 364, 515

\end{thebibliography}
\end{document}